\documentclass[journal,12pt,onecolumn,draftclsnofoot,]{IEEEtran}
\usepackage{graphicx}
\usepackage{subcaption}
\usepackage{cite}
\usepackage{multirow}
\usepackage{amssymb}
\usepackage{scalerel}
\usepackage{algorithm}
\usepackage{algpseudocode}
\normalsize

\usepackage{amsthm}

\usepackage[cmex10]{amsmath}
\usepackage [english]{babel}
\usepackage [autostyle, english = american]{csquotes}
\MakeOuterQuote{"}

\begin{document}

%
\title{Energy Efficient Power Allocation for Device-to-Device Communications Underlaid  Cellular Networks Using Stochastic Geometry}

\author{Negar~Zabetian,
	Abbas~Mohammadi,
	and~Meysam~Masoudi
\thanks{N. Zabetian and A. Mohammadi are with the Department of Electrical Engineering, Amirkabir University of Technology (Tehran Polytechnic), Tehran, Iran.
	
	 Meysam Masoudi is with the School of Electrical Engineering and Computer Science, KTH Royal Institute of Technology, Stockholm, Sweden}}

%



\maketitle
\begin{abstract}
In this paper, we study an energy efficiency maximization problem in uplink for D2D communications underlaid with cellular networks on multiple bands. Utilizing stochastic geometry, we derive closed-form expressions for the average sum rate, successful transmission probability, and energy efficiency of cellular and D2D users. Then, we formulate an optimization problem to jointly maximize the energy efficiency of D2D and cellular users and obtain optimum transmission power of both D2D and cellular users. In the optimization problem,  we guarantee the QoS of users by taking into account the success transmission probability on each link.  To solve the problem, first we convert the problem into canonical convex form. Afterwards, we solve the problem in two phases, energy efficiency maximization of devices and energy efficiency maximization of cellular users. In the first phase, we maximize the energy efficiency of D2D users and feed the solution to the second phase where we maximize the energy efficiency of cellular users. Simulation results reveal that significant energy efficiency can be attained  e.g., 10\% energy efficiency improvement compared to fix transmission power in high density scenario.
\end{abstract}


\IEEEpeerreviewmaketitle



%

\section{Introduction}

\IEEEPARstart With the unprecedented demand of mobile traffic, the fifth generation (5G) of mobile networks are anticipated to support 1000 times more data traffic \cite{zhou2016energy}. This requirement urges the development of new technologies, and among them, device-to-device (D2D) communication is considered as one of the key technologies that aid 5G networks to achieve their goals. In D2D communication, proximate devices can directly communicate with each other under the control of base stations (BSs) \cite{2}. Due to the proximity of devices, D2D communication has the advantage of providing services with higher data rate,  spectral efficiency (SE),  energy efficiency (EE) with lower power consumption and latency \cite{3}. Despite the appealing potential gains of D2D communication, sharing the same spectrum with cellular users induce new challenges into the network such as interference management. D2D communications can operate on either licensed band (in-band) or unlicensed band (out-band). In the literature, in-band communication is classified into two categories, namely, underlay and overlay communications. In the overlay scenario, the system reserves dedicated parts of the spectrum for D2D links and cellular users use another part of the spectrum to avoid interference between D2D users and cellular users. However, in this case, SE is less than underlay mode since the spectrum is underutilized. In the underlay communication, D2D devices reuse the cellular spectrum and hence create interference on cellular users \cite{4}. Therefore, there is a need to utilize efficient resource allocation approaches to take advantage of D2D benefits and mitigate the challenges of using such technology.

In the literature, there are bulks of studies investigating the efficient resource allocation to benefit from the advantages of D2D communications while diminishing the performance degradation of the cellular system due to new sources of interference. In particular, power control is one of the promising approaches that mitigates interference \cite{lee2015power}. In  \cite{6} a power control scheme is proposed to manage the cross-tier interference between cellular and D2D users.  The authors in \cite{zhang2013interference} proposed an interference-aware algorithm for power control in D2D communications underlaying cellular networks. In \cite{memmi2017power}, a single-cell D2D underlay cellular network is considered. A centralized and distributed algorithms are proposed to find the users' optimal power transmission.   A dynamic power control for D2D communication underlaying uplink multi-cell network is investigated in \cite{jiang2018resource} considering interference mitigation. However power control is a key technique to manage the interference, both the energy efficiency (EE) and QoS are still influenced by the interference in the network.

Energy efficiency (EE) is a metric quantifying  the  efficiency of resource utilization. In fact, EE not only brings considerable economic benefits into the network but also can be interpreted as  concerning about the environment \cite{masoudi2017energy}. Extensive research studies have been devoted to the energy efficiency of the system.  EE maximization in D2D communication underlaying cellular network in cloud radio access network is studied in \cite{zhou2016energy}. A closed-form expression for EE is derived in \cite{11} by utilizing stochastic geometry tool. In this study, EE is maximized by optimizing transmit power and density of base station (BS) in cellular networks. An energy efficient power control algorithm is proposed in \cite{wu2015optimal} to share resources among the cellular and D2D users. Joint power and subcarrier  allocation is proposed in  \cite{12,13} for energy efficient D2D communication  on  multiple bands and single-cell scenario. EE and SE trade-off  in a single-cell  D2D communication scenario  is investigated in \cite{14} by a distributed resource allocation scheme.  The optimal power control for energy efficiency of D2D communication underlaying cellular networks is analyzed in \cite{yang2016optimal} where only a single band is considered. In \cite{bhardwaj2018energy}, EE maximization problem with a constraint on SE is investigated in D2D communication underlay cellular network on multiple bands. A power allocation algorithm is proposed to obtain optimum power of D2D users. However, they did not consider the  power allocation for cellular users.  The authors in \cite{hoang2016energy} formulated an energy-efficient resource allocation problem considering multiple bands; however the QoS is limited to the minimum required rate for cellular users.  The D2D communication underlaying cellular networks on multiple bands is considered in \cite{15}, and the derivative-based algorithm is applied to maximize the energy efficiency of D2D users, however they did not optimize the cellular users transmission power. In this study, the authors derived the outage transmission probability as QoS metric using Stochastic geometry. 

Stochastic geometry is a powerful tool to model and analyze D2D networks and provides closed-form expressions for several metrics such as average sum rate and successful transmission probability \cite{8,9,abdallah2017distance}. In \cite{8}, the authors model the D2D users distribution by Poisson point process (PPP).  In \cite{9}, the outage probability for a cellular user and a D2D receiver is obtained. In \cite{abdallah2017distance}, a power control scheme is proposed to improve the outage probability in the network.

According to the best of our knowledge, there are few studies in the literature obtaining optimum transmission power of D2D users in the cellular networks where users are modeled based on stochastic geometry on multiple bands. The authors in \cite{15} modeled the users by stochastic geometry on multiple bands and maximized the EE of D2D users. However, they neglect the impact of  cellular users' transmission power.
In this paper, we bring up an effective solution to maximize the energy efficiency of cellular network underlaid D2D communication in multiple bands and derived optimum power in each band. Because the channel fading coefficient can be varying in different bands. Also, there are different densities for D2D and cellular users in each band and the number of users which are allowed to transmit is different. Due to the fluctuation of resources and the evolving radio and network conditions,  having multiple bands is more practical in future 5G wireless networks, and it can consider the dynamic behavior of a system.
Since the wireless channel parameters such as channel fading may vary, having multiple bands and adjustable power transmission on each, can enhance the energy efficiency of the network. Moreover, due to having multiple bands, D2D users do not interfere on the cellular users and devices in other bands, so, the complexity of interference management becomes affordable, and the performance of D2D communications is more acceptable. 

We derived the closed-form expressions for the average sum rate (ASR) and successful transmission probability (STP) for both the cellular users and D2D users. Then, we formulated a mixed integer nonlinear programming (MINLP) optimization problem to maximize the energy efficiency of both cellular and D2D users while considering the QoS  of both cellular and D2D users. To solve the problem, first, we transformed the MINLP problem into a convex form utilizing the technique of changing the optimization variable. We have also broken down the problem into two sub-problems. To address the primary problem, we solve these sub-problems iteratively. The first sub-problem maximizes the EE of D2D users and finds the optimal transmission power  of D2D users. By using the solution of the first sub-problem, the second problem finds the optimal transmission power of cellular users to maximize the EE.  As a reference case, we compared our study with \cite{15} which has similar system model and assumptions to our paper.

The remainder of the paper is organized as follows: The system model is presented in Section II. In Section III, the analytic studies and problem formulations are described. The solution methodology is proposed in Section IV followed by Section V where the numerical results are discussed. Finally, in Section VI the concluding remarks are presented.

Notation: $\mathbb{P(.)}$, $\mathbb{E}(x)$ and $\exp(.)$ represent the probability, expectation value of random variable $x$ and exponential function, respectively. $\rm\Gamma(.)$ expresses  the gamma function with the form $\rm\Gamma (z) = \int\limits_0^\infty  {{t^{z - 1}}{e^{ - t}}} dt$ and ${\mathcal{L}_{x}}\left( s \right)$ stands for the Laplace transform of $x$.

\section{system model}
\begin{figure}[ht]
	\centering
	\includegraphics[width=0.5\linewidth]{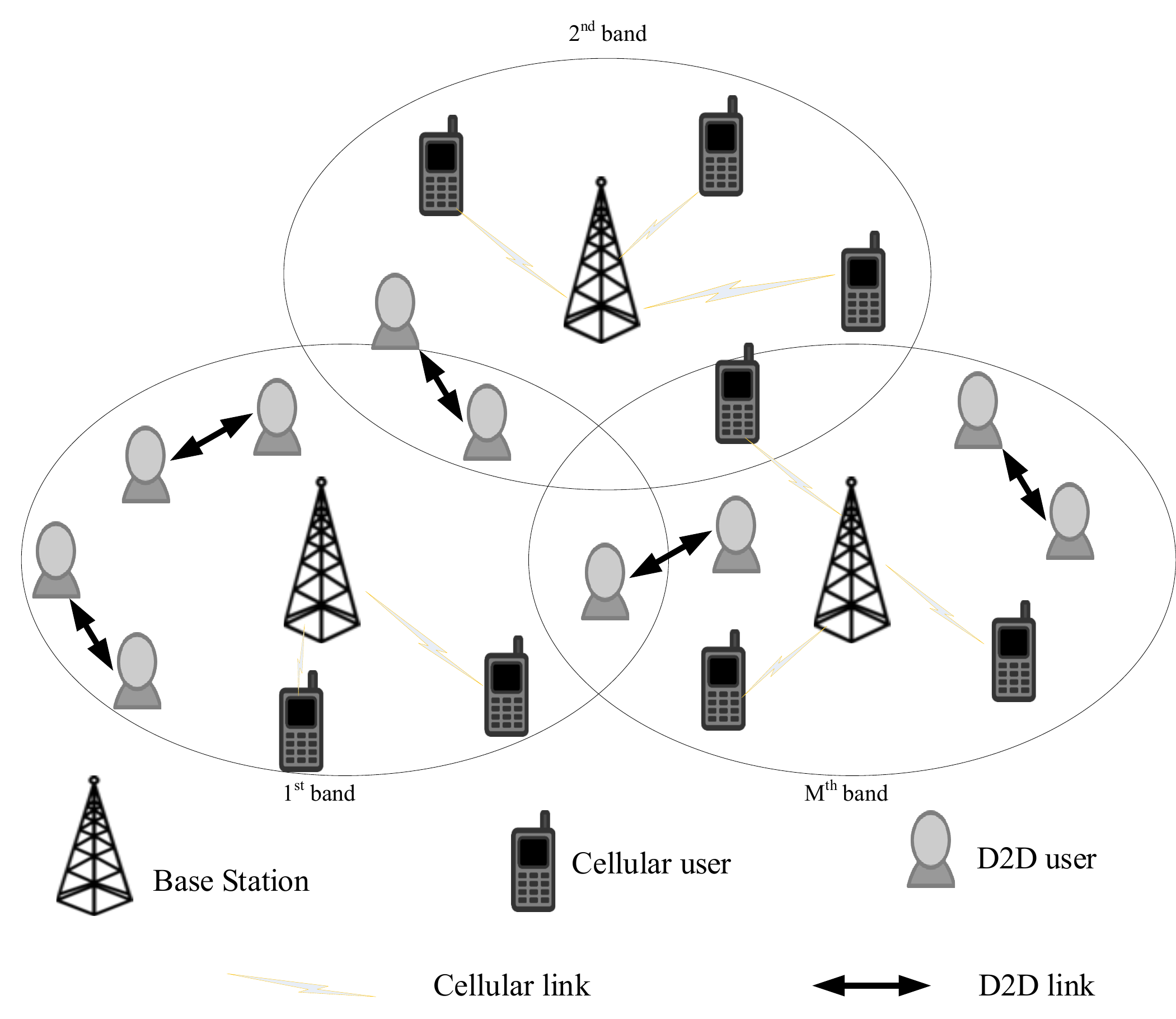}
	\caption{A representation of D2D communication underlaying cellular network. 
	} \label{fig:sys}
\end{figure}

In this study, we investigate a cellular network underlaid with D2D communications in the uplink direction as depicted in Fig. \ref{fig:sys}. In this system, the set of active cellular users in each band share the same bandwidth with the set of active D2D users in that band. Cellular and D2D users are distributed over the network with homogeneous PPP distribution. As can be seen in Fig. \ref{fig:sys}, all cellular users and devices are distributed on a two-dimensional plane $\mathbb{R}^2$.  In the rest of this paper, superscript C and D are used to denote cellular and D2D users. We denote the total bandwidth by $W$, and we break it down to $M$ sub-bands. The bandwidth of each sub-band is $W_{i}$ where  $i=1, 2, \dots, M$.  
Since the wireless channel parameters such as channel fading may vary, having multiple bands and adjustable power transmission on each, can enhance the energy efficiency of the network. We assume that the signals in different bands do not interfere with each other. Both cellular and D2D users can utilize these sub-bands  with density ${\lambda _{c,i}}$ and ${\lambda _{d,i}}$ on each band, respectively. In each band, there are different densities for D2D and cellular users. If the more bandwidth is allocated to the users, the more users are allowed to transmit, and the density of users become greater. We consider that the channel model has path loss and  small-scale fading which is modeled by Rayleigh distribution with unit mean.  Therefore, channel gain between users has an exponential distribution which is denoted by $g$.
The received power of both devices and cellular users are obtained by considering large- and small scale fading as follows. 
\begin{equation}
{P_r} = {P_t}g{r^{ - \alpha }},
\end{equation}
where $\alpha$ is path loss exponent, $r$ is the distance between receiver and transmitter and $P_t$ is the transmission power.

According to Palm theorem \cite{16}, for Poisson point process, each point has a similar probability of being chosen. Therefore, we have  \textit{randomly chosen}  point or \textit{typical}  point. According to Slivnyak's theorem \cite{16}, conditioning on a point does not influence the distribution of the rest of the process. Thus, we can condition on having a typical D2D receiver and a typical base station for D2D and cellular communication at the center on $\mathbb{R}^2$, respectively and analyze the performance of the whole network.

The users on the same band can interfere with each other while they have no interference on other bands. Assume the  signal which is received at the D2D  receiver in $i$-th band is denoted by $Y_{d,i}$ and calculated as, 

\begin{equation}\label{eq:1}
\begin{array}{*{20}{l}}
{{Y_{d,i}} = \overbrace {{s_{d,i}}{h_{d,00}}R_{d,00,i}^{ - \frac{\alpha }{2}}}^{Signal{\kern 1pt} from{\kern 1pt} typical{\kern 1pt} D2D{\kern 1pt} user} + \underbrace {\sum\limits_{j \in {\phi _{c,i}}} {{s_{c,i}}{h_{c,j0}}R_{c,j0,i}^{ - \frac{\alpha }{2}}} }_{Interference{\kern 1pt} from{\kern 1pt} cellular{\kern 1pt} users}}
{+ \overbrace {\sum\limits_{\ell  \in {\phi _{d,i}}} {{s_{d,i}}{h_{d,\ell 0}}R_{d,\ell 0,i}^{ - \frac{\alpha }{2}}} }^{Interference{\kern 1pt} from{\kern 1pt} D2D{\kern 1pt} users} + n},
\end{array}
\end{equation}
where ${\phi _{d,i}}$ and ${\phi _{c,i}}$ denote the set of active D2D and cellular users in $i$-th band, ${s_{d,i}}$ and ${s_{c,i}}$ represent information signals of D2D and cellular receiver in $i$-th band, respectively.  ${h_{d,00}}$ and ${R_{d,00,i}}$ are the channel coefficient and the distance between the typical D2D receiver and the corresponding transmitter in the $i$-th band, respectively. ${h_{c,j0}}$ and ${R_{c,j0,i}}$ represent the channel coefficient and the distance between the $j$-th cellular transmitter and the typical D2D receiver in the $i$-th band, respectively. Similarly, ${h_{d,\ell 0}}$ and ${R_{d,\ell 0,i}}$ are the channel coefficient and the distance between the $\ell$-th D2D transmitter and the typical D2D receiver in the $i$-th band, respectively.

$n$ is additive white Gaussian noise (AWGN) with zero mean and variance $N_{0}$. The signal to noise plus interference ratio (SINR) of typical D2D receiver in $i$-th band is given by 
\begin{equation}\label{eq:2}
SIN{R_{d,i}} =  \frac {{{g_{d,00}}R_{d,00,i}^{ - \alpha }}} {{{I_{d,c0,i}} + {I_{d,d0,i}} +   \frac{N_0}{{P_{d,i}}} }}.
\end{equation}
The interference which is introduced by cellular users to the typical D2D receiver is given by
\begin{equation}
{I_{d,c0,i}} = \sum\limits_{j \in {\Phi _{c,i}}} {\frac{{{P_{c,i}}}}{{{P_{d,i}}}}{g_{c,j0}}R_{c,j0,i}^{ - \alpha }} ,
\end{equation}
where ${P_{d,i}}$  and ${P_{c,i}}$ are transmission power of the D2D and cellular users in the $i$-th band.

Also, the interference from D2D users to the typical D2D receiver in the $i$-th band is given by
\begin{equation}
{I_{d,d0,i}} =\sum\limits_{\ell  \in {\Phi _{d,i}}} {{g_{d,\ell 0}}R_{d,\ell 0,i}^{ - \alpha }}. 
\end{equation}
Similarly, the received signal by the typical base station is given by 
\begin{equation}\label{eq:3}
\begin{array}{l}
{Y_{c,i}} = {s_{c,i}}{h_{c,00}}R_{c,00,i}^{ - \frac{\alpha }{2}} + \sum\limits_{j \in {\phi _{c,i}}} {{s_{c,i}}{h_{c,j0}}R_{c,j0,i}^{ - \frac{\alpha }{2}}} 
\begin{array}{*{20}{c}}
{}&{+ \sum\limits_{\ell  \in {\phi _{d,i}}} {{s_{d,i}}{h_{d,\ell 0}}R_{d,\ell 0,i}^{ - \frac{\alpha }{2}}}  + n.}
\end{array}
\end{array}
\end{equation}
The SINR of the typical BS in $i$-th band is as follows. 
\begin{equation}\label{eq:4}
SIN{R_{c,i}} = \frac {{ {g_{c,00}}R_{c,00,i}^{ - \alpha }}} {{{I_{c,c0,i}} + {I_{c,d0,i}} + \frac{N_0}{{P_{c,i}}} }},
\end{equation}
The interference from cellular users to the typical BS in $i$-th band is shown by
\begin{equation}
{I_{c,c0,i}} = \sum\limits_{j \in {\Phi _{c,i}}} {{g_{c,j0}}R_{c,j0,i}^{ - \alpha }}.
\end{equation}
Also, the interference from  D2D users to the typical BS in $i$-th band is shown by
\begin{equation}
{I_{c,d0,i}} = \sum\limits_{\ell  \in {\Phi _{d,i}}} {\frac{{{P_{d,i}}}}{{{P_{c,i}}}}{g_{d,\ell 0}}R_{d,\ell 0,i}^{ - \alpha }} .
\end{equation}

Similar to other studies, i.e., \cite{8}, \cite{15}, in the rest of this study, the performance of the system  is studied  under the interference-limited regime,  where the interference, which is introduced by spectrum sharing between large number of users, is dominating the noise power.  Hence,   the impact of thermal noise can be neglected and the SINR expression is converted to signal to interference ratio (SIR).

\section{Problem Formulation}
In this section, we intend to maximize the energy efficiency of the network. To achieve this goal, first, we introduce the total energy efficiency (EE) as the ratio of average sum rate (ASR) to the total power consumption \cite{15}. Then, we use the closed-form expressions for the ASR using the success probability formula in \cite{15}. Afterwards, we formulate the energy efficiency maximization as an optimization problem. 

\subsection{Users Success Probability}
Total energy efficiency of the network is defined as, 
\begin{equation}
EE = \frac{{ASR}}{P_t},
\end{equation}
where $ASR$ and $P_t$ are average sum rate and total transmission power of all users. To calculate $P_t$, the power consumption per unit area in $i$-th band is considered which is expressed as ${\lambda _{d,i}}{P_{d,i}}$ and ${\lambda _{c,i}}{P_{c,i}}$ for D2D and cellular users, respectively.  $ASR_{d,i}$ is defined as ${\lambda _{d,i}}{\bar R_{d,i}}$. To compute ASR, we use the lower bounds on the average rates of cellular and D2D users \cite{17}.
\begin{equation}\label{eq:5}
\begin{array}{l}
{{\bar R}_{c,i}} = {\rm \mathbb{E}}\left[{W_i} {{{\log }_2}\left( {1 + SIR_{c,i}} \right)} \right] = \mathop {\sup }\limits_{{T_{c,i}} \ge 0} {W_i}\log_2(1 + {T_{c,i}})\mathbb{P} (SI{R_{c,i}} \ge {T_{c,i}}),\\
{{\bar R}_{d,i}} = {\rm \mathbb{E}}\left[{W_i} {{{\log }_2}\left( {1 + SIR_{d,i}} \right)} \right] = \mathop {\sup }\limits_{{T_{d,i}} \ge 0} {W_i}\log_2(1 + {T_{d,i}})\mathbb{P} (SI{R_{d,i}} \ge {T_{d,i}}),
\end{array}
\end{equation}
where ${T_{c,i}}$ and ${T_{d,i}}$ are SIR thresholds for cellular and D2D users in $i$-th band.

\textit{Theorem 1.} The successful transmission probability for typical D2D receiver in $i$-th band is given by
\begin{equation}\label{eq:6}
\mathbb{P} (SI{R_{d,i}} \ge {T_{d,i}}) = \exp \bigg\{ {\bigg. { - {\varsigma _{d,i}}\Big[ {{\lambda _{d,i}} + {\lambda _{c,i}}{{\big( {\frac{{{P_{c,i}}}}{{{P_{d,i}}}}} \big)}^{\frac{2}{\alpha }}}} \Big]} \bigg\}} \bigg. , 
\end{equation}
where ${\lambda _{d,i}}$  and ${\lambda _{c,i}}$  are the density of D2D and cellular users in $i$-th band, respectively. Moreover, ${\varsigma _{d,i}} \buildrel \Delta \over = \pi {T_{d,i}}^{\frac{2}{\alpha }}R_{d,00,i}^2\rm\Gamma \big( {1 + \frac{2}{\alpha }} \big)\rm\Gamma \big( {1 - \frac{2}{\alpha }} \big)$ \cite{15}.
\begin{proof}
	By using (\ref{eq:2}), we have
	\begin{equation}\label{eq:7}
	\begin{array}{l}
	\mathbb{P} \big(SI{R_{d,i}} \ge {T_{d,i}}\big)\\
	= \mathbb{P} \big( {\frac{{ {g_{d,00}}R_{d,00,i}^{ - \alpha }}}{{{I_{d,c0,i}} + {I_{d,d0,i}}}} \ge {T_{d,i}}} \big)\\
	= \mathbb{P} \Big( {{g_{d,00}} \ge {T_{d,i}}R_{d,00,i}^\alpha \big( {{I_{d,c0,i}} + {I_{d,d0,i}}} \big)} \Big)\\
	\mathop  = \limits^{(a)} \mathbb{E}\Big( {\prod\limits_{\ell \in {\phi _{d,i}}} {\exp \big( { - {T_{d,i}}R_{d,00,i}^\alpha {g_{d,\ell 0}}R_{d,\ell 0,i}^{ - \alpha }} \big)} } \Big)
	\times \mathbb{E}\Big( {\prod\limits_{j  \in {\phi _{c,i}}} {\exp \big( { - {T_{d,i}}R_{d,00,i}^\alpha \frac{{{P_{c,i}}}}{{{P_{d,i}}}}{g_{c,j0}}R_{c,j0,i}^{ - \alpha }} \big)} } \Big)\\
	= {\mathcal{L}_{{I_{d,d0,i}}({g_{d,\ell 0}})}}({T_{d,i}}R_{d,00,i}^\alpha ){\mathcal{L}_{{I_{d,c0,i}}({g_{c,j0}})}}({T_{d,i}}R_{d,00,i}^\alpha ),
	\end{array}
	\end{equation}
	where (a) is due to exponential distribution and independence of cellular and D2D channel gains.
	Considering Laplace transform and stochastic geometry theorem \cite{16}, we have  
	\begin{equation}\label{eq:8}
	\begin{array}{l}
	{\mathcal{L}_{{I_{d,d0,i}}({g_{d,\ell 0}})}}({T_{d,i}}R_{d,00,i}^\alpha )
	= \exp \Big[ { - {\lambda _{d,i}}\int\limits_0^\infty  {E\big( {{g_{d,\ell 0}}} \big)} \big( {1 - {e^{ - {T_{d,i}}R_{d,00,i}^\alpha {r^{ - \alpha }}}}} \big)dr} \Big]
	\\\,\,\,\,\,\,\,\,\,\,\,\,\,\,\,\,\,\,\,\,\,\,\,\,\,\,\,\,\,\,\,\,\,\,\,\,\,\,\,\,\,\,\,\,\,\,\,\,\,\,\,\,\,\,\,\,\,=  \exp \Big[ { - {\lambda _{d,i}}\pi T_{d,i}^{\frac{2}{\alpha }}R_{d,00,i}^2\rm\Gamma \big( {1 + \frac{2}{\alpha }} \big)\rm\Gamma \big( {1 - \frac{2}{\alpha }} \big)} \Big],\\
	{\mathcal{L}_{{I_{d,c0,i}}({g_{c,j0}})}}({T_{d,i}}R_{d,00,i}^\alpha )
	= \exp \Big[ { - {\lambda _{c,i}}{{\left( {\frac{{{P_{c,i}}}}{{{P_{d,i}}}}} \right)}^{\frac{2}{\alpha }}}\pi T_{d,i}^{\frac{2}{\alpha }}R_{d,00,i}^2\rm\Gamma \big( {1 + \frac{2}{\alpha }} \big)\rm\Gamma \big( {1 - \frac{2}{\alpha }} \big)} \Big], 
	\end{array}
	\end{equation}
	
	Substituting (\ref{eq:8}) in (\ref{eq:7}) concludes the proof.   
\end{proof}

\textit{Theorem 2.} The STP for typical base station in $i$-th band is obtained as,
\begin{equation}\label{eq:9}
\mathbb{P} (SI{R_{c,i}} \ge {T_{c,i}}) = \exp \bigg\{ {\bigg. { - {\varsigma _{c,i}}\Big[ {{\lambda _{c,i}} + {\lambda _{d,i}}{{\big( {\frac{{{P_{d,i}}}}{{{P_{c,i}}}}} \big)}^{\frac{2}{\alpha }}}} \Big]} \bigg\}} \bigg. ,
\end{equation}
where ${\varsigma _{c,i}}  \buildrel \Delta \over =  \pi {T_{c,i}}^{\frac{2}{\alpha }}R_{c,00,i}^2\rm\Gamma \big({1 + \frac{2}{\alpha }} \big)\rm\Gamma \big( {1 - \frac{2}{\alpha }} \big)$. 
\begin{proof}
	The proof is obtained similarly to Theorem 1 \cite{15}. 
\end{proof}

By substituting, (\ref{eq:6}) and (\ref{eq:9}) in (\ref{eq:5}) ASR of cellular users and devices are obtained. Therefore, EE of cellular and D2D users in $i$-th band are given by,
\begin{equation}\label{eq:10}
\begin{array}{l}
E{E_{c,i}} = \frac{{AS{R_{c,i}}}}{{{\lambda _{c,i}}{P_{c,i}}}}= \frac{{{W_i}}}{{{P_{c,i}}}}\log_2\big(1 + {T_{c,i}}\big)\exp \bigg\{ {\left. { - {\varsigma _{c,i}}\Big[ {{\lambda _{c,i}} + {\lambda _{d,i}}{{\big( {\frac{{{P_{d,i}}}}{{{P_{c,i}}}}} \big)}^{\frac{2}{\alpha }}}} \Big]} \right\}} \bigg.,\\
E{E_{d,i}} = \frac{{AS{R_{d,i}}}}{{{\lambda _{d,i}}{P_{d,i}}}}= \frac{{{W_i}}}{{{P_{d,i}}}}\log_2\big(1 + {T_{d,i}}\big)\exp \bigg\{ {\left. { - {\varsigma _{d,i}}\Big[ {{\lambda _{d,i}} + {\lambda _{c,i}}{{\big( {\frac{{{P_{c,i}}}}{{{P_{d,i}}}}} \big)}^{\frac{2}{\alpha }}}} \Big]} \right\}} \bigg.
\end{array}
\end{equation}

\subsection{Energy Efficiency Optimization Problem}
In the previous subsection, we defined energy efficiency based on the success probability of cellular and D2D communications. Since the interference between cellular users and devices, has an impact on the energy efficiency of each user, it is more practical to study the EE of the  entire cellular network. In what follows, we will define energy efficiency maximization problems for D2D and cellular users, separately and find optimal power of D2D and cellular users.

\subsubsection{D2D Users Energy Efficiency}

In this subsection, we formulate an optimization problem to maximize the total energy efficiency of D2D users. EE optimization problem for D2D users is expressed as follows.
\begin{equation}\label{eq:11}
\begin{array}{l}
\mathop {\max }\limits_{{P_{d,i}}} \begin{array}{*{20}{c}}
{}&{E{E_d}}
\end{array} = \sum\limits_{i = 1}^M {E{E_{d,i}}} \\
s \cdot t \cdot \,\,\,\,\,\,(1)\,\sum\limits_{i = 1}^M {{P_{d,i}} \le {P_d}} \\
\begin{array}{*{20}{c}}
{}&{\,\,\,\,\,\,\,\,\,\,\,\,(2)\,0 \le {P_{d,i}} \le {P_{d,i,\max }}}
\end{array}\\
\begin{array}{*{20}{c}}
{}&{\,\,\,\,\,\,\,\,\,\,\,\,(3)\,1 - \mathbb{P}(SI{R_{c,i}} \ge {T_{c,i}})}
\end{array} \le {\theta _{c,i}}\\
\begin{array}{*{20}{c}}
{}&{\,\,\,\,\,\,\,\,\,\,\,\,(4)\,1 -  \mathbb{P}(SI{R_{d,i}} \ge {T_{d,i}})}
\end{array} \le {\theta _{d,i}},
\end{array}
\end{equation}
In (\ref{eq:11}), $EE_{d,i}$  is the energy efficiency of  D2D users in band $i$. 
The first constraint ensures that the total transmission powers of D2D users over all bands do not exceed the maximum possible value which is denoted by ${P_d}$. The second constraint ensures that the power of D2D users in each band cannot exceed an allowable threshold indicated by ${P_{d,i,\max }}$.  To satisfy the  quality of service of users, the outage probability of both cellular and D2D users should be less than predefined thresholds which are denoted by ${\theta _{c,i}}$  and  ${\theta _{d,i}}$, respectively. According to the quality of service constraints and objective function, the problem is not convex.

\subsubsection{Cellular Users Energy Efficiency}
In the previous subsection, when maximizing energy efficiency, we assumed a constant power for cellular users. Since cellular users are operating in the same band as D2D users, there is more room to enhance the energy efficiency of the entire network by optimizing the power consumption of cellular users. In this subsection, we calculate the optimal transmission power of cellular users in each band. 
Energy efficiency optimization problem for cellular users is formulated as follows.
\begin{equation} \label{eq:cell_opt_prob}
\begin{array}{l}
\mathop {\max }\limits_{{P_{c,i}}} \begin{array}{*{20}{c}}
{}&{E{E_c}}
\end{array} = \sum\limits_{i = 1}^M {E{E_{c,i}}} \\
s \cdot t \cdot \,\,\,\,\,\,(1)\,\sum\limits_{i = 1}^M {{P_{c,i}} \le {P_c}} \\
\begin{array}{*{20}{c}}
{}&{\,\,\,\,\,\,\,\,\,\,\,\,(2)\,0 \le {P_{c,i}} \le {P_{c,i,\max }}}
\end{array}\\
\begin{array}{*{20}{c}}
{}&{\,\,\,\,\,\,\,\,\,\,\,\,(3)\,\sum\limits_{i = 1}^M {{P_{d,i}} \le {P_d}}}
\end{array}\\
\begin{array}{*{20}{c}}
{}&{\,\,\,\,\,\,\,\,\,\,\,\,(4)\,0 \le {P_{d,i}} \le {P_{d,i,\max }}}
\end{array}\\
\begin{array}{*{20}{c}}
{}&{\,\,\,\,\,\,\,\,\,\,\,\,(5)\,1 - \mathbb{P}(SI{R_{c,i}} \ge {T_{c,i}})}
\end{array} \le {\theta _{c,i}}\\
\begin{array}{*{20}{c}}
{}&{\,\,\,\,\,\,\,\,\,\,\,\,(6)\,1 -  \mathbb{P}(SI{R_{d,i}} \ge {T_{d,i}})}
\end{array} \le {\theta _{d,i}},
\end{array}
\end{equation}
where $EE_{c,i}$  is the energy efficiency of  cellular users in band $i$. The total transmission power of cellular users over all bands and transmission power of cellular users in each band should be less than predefined thresholds which are denoted by ${P_c}$  and ${P_{c,i,\max }}$, respectively. Moreover, D2D users consume more power to coordinate the interference which is produced by cellular communication. Thus, we have constraints for the power of devices besides cellular users' power constraints. Also, we have the quality of service constraints similar to the optimization problem of D2D users.
\section{Solution Methodology}
In this section, we aim at addressing the optimization problem. First, we transform the optimization problems into the canonical convex form by a technique of changing the variable. Then, we solve the problem of  D2D  and then the corresponding results are given as an input to the cellular users' optimization problem. 

\subsection{D2D Users Energy Efficiency}
To transform the problem defined in (\ref{eq:11}), we define a new optimization variable, ${x_i}$, as, 
\begin{eqnarray} \label{eq:change}
{x_i} \buildrel \Delta \over = \exp \Big( {{\varsigma _{d,i}}{\lambda _{c,i}}{{\big( {\frac{{{P_{c,i}}}}{{{P_{d,i}}}}} \big)}^{\frac{2}{\alpha }}}} \Big).
\end{eqnarray}

Therefore, by inserting $x_i$ into (\ref{eq:6}), (\ref{eq:9}) and (\ref{eq:10}), the optimization problem defined in (\ref{eq:11})  can be rewritten as follows. 
\begin{equation}\label{eq:12}
\begin{array}{l}
\mathop {\max }\limits_{{x_i}} \,\,\,E{E_d} = \sum\limits_i {\frac{{{W_i}\log_2\big(1 + {T_{d,i}}\big)\exp \big( { - {\varsigma _{d,i}}{\lambda _{d,i}}} \big){{\big( {\ln{x_i}} \big)}^{^{\frac{\alpha }{2}}}}}}{{{P_{c,i}}{x_i}{{\big( {{\varsigma _{d,i}}{\lambda _{c,i}}} \big)}^{^{\frac{\alpha }{2}}}}}}} \\
s \cdot t \cdot \,(1)\,\sum\limits_i {{P_{c,i}}{{\left( {\frac{{{\varsigma _{d,i}}{\lambda _{c,i}}}}{{\ln{x_i}}}} \right)}^{^{\frac{\alpha }{2}}}}}  \le {P_d}\\
\,\,\,\,\,\,\,\,\,\,\,\,\,\,\,(2)\,{P_{c,i}}{\big( {\frac{{{\varsigma _{d,i}}{\lambda _{c,i}}}}{{\ln{x_i}}}} \big)^{^{\frac{\alpha }{2}}}} \le {P_{d,i,\max }}\\
\begin{array}{*{20}{c}}
{}&{\,\,\,\,\,\,\,\,(3)\,}
\end{array}{x_i} \le \frac{{\exp \big( { - {\varsigma _{d,i}}{\lambda _{d,i}}} \big)}}{{1 - {\theta _{d,i}}}}\\
\begin{array}{*{20}{c}}
{}&{\,\,\,\,\,\,\,\,(4)}
\end{array}{x_i} \ge \exp \big( {\frac{{ - {\varsigma _{d,i}}{\lambda _{c,i}}}}{{\frac{{ln(1 - {\theta _{c,i}})}}{{{\varsigma _{c,i}}{\lambda _{d,i}}}} + \frac{{{\lambda _{c,i}}}}{{{\lambda _{d,i}}}}}}} \big).  
\end{array}
\end{equation}
\textit{Theorem 3.} $E{E_{d,i}}$ is concave on the interval $({t_1},{t_2})$ and convex on the interval $(1,{t_1}) \cup ({t_2},+\infty )$,
where ${t_{1,2}} \buildrel \Delta \over = \exp (\frac{{3\alpha  \mp \sqrt {{\alpha ^2} + 16\alpha } }}{8})$.
\begin{proof}
	If the second derivative is non-positive or non-negative, the function is concave or convex  \cite{18}. We have
	\begin{equation}\label{eq:13}
	\begin{array}{l}
	\frac{{{\partial ^2}E{E_{d,i}}}}{{\partial x_i^2}}
	= {A_i}\frac{{2{{\left( {\ln{x_i}} \right)}^{\frac{\alpha }{2}}} - \frac{{3\alpha }}{2}{{\left( {\ln{x_i}} \right)}^{\frac{\alpha }{2} - 1}} + \left( {\frac{\alpha }{2} - 1} \right)\frac{\alpha }{2}{{\left( {\ln{x_i}} \right)}^{\frac{\alpha }{2} - 2}}}}{{{P_{c,i}}x_i^3}},
	\end{array}
	\end{equation}
	where ${A_i} \buildrel \Delta \over = \frac{{{W_i}\log_2(1 + {T_{d,i}})\exp \left( { - {\varsigma _{d,i}}{\lambda _{d,i}}} \right)}}{{{{\left( {{\varsigma _{d,i}}{\lambda _{c,i}}} \right)}^{\frac{\alpha }{2}}}}}$.
	$t_1$ and $t_2$ are the solution of $\frac{{{\partial ^2}E{E_{d,i}}}}{{\partial x_i^2}} = 0$. The second derivative is positive on the interval $(1,{t_1}) \cup ({t_2},+\infty )$ and negative on the interval $({t_1},{t_2})$. Hence $E{E_{d,i}}$ is convex in the former and  concave in the latter interval. This completes the proof. 
\end{proof}

The objective function is the summation of concave (convex) functions in defined sets,  and the problem is solved in the intersection of these feasible sets. Therefore, the domain of the objective function is the intersection between the domain of each ${EE_{d,i}}$. Also the constraints are convex. Since the second derivative of constraints (1) and (2) are always positive, and the constraints (3) and (4) are linear functions of ${x_i}$. Consequently, the optimization problem is convex \cite{18}. Therefore, to find the optimal points, the problem is solved in the concave interval.

\subsection{Cellular Users Energy Efficiency}

Similar to the approach explained for D2D scenario, by substituting $x_i$ in (\ref{eq:10}), the optimization problem defined in (\ref{eq:cell_opt_prob}) can be rewritten as, 

\begin{equation}\label{eq:14}
\begin{array}{l}
\mathop {\max }\limits_{{P_{c,i}}} \,\,E{E_c} = \sum\limits_{i = 1}^M {\frac{{{W_i}\log_2(1 + {T_{c,i}})\exp \left( { - {\varsigma _{c,i}}{\lambda _{c,i}}} \right)}}{{{P_{c,i}}\exp \left( {\frac{{{\varsigma _{d,i}}{\varsigma _{c,i}}{\lambda _{d,i}}{\lambda _{c,i}}}}{{\ln{x_i}}}} \right)}}} \\
s \cdot t \cdot \,\,\,\,\,\,\,\,\,\,(1)\,\sum\limits_i {{P_{c,i}}}  \le {P_c}\\
\begin{array}{*{20}{c}}
{}&{}
\end{array}\,\,\,\,\,\,\,\,\,\,\,\,\,(2)\,0 \le {P_{c,i}} \le {P_{c,i,\max }}\\
\begin{array}{*{20}{c}}
{}&{}
\end{array}\,\,\,\,\,\,\,\,\,\,\,\,\,(3)\,\sum\limits_i {{P_{c,i}}{{\left( {\frac{{{\varsigma _{d,i}}{\lambda _{c,i}}}}{{\ln{x_i}}}} \right)}^{\frac{\alpha }{2}}}}  \le {P_d}\\
\begin{array}{*{20}{c}}
{}&{}
\end{array}\,\,\,\,\,\,\,\,\,\,\,\,\,(4)\,{P_{c,i}}{\left( {\frac{{{\varsigma _{d,i}}{\lambda _{c,i}}}}{{\ln{x_i}}}} \right)^{\frac{\alpha }{2}}} \le {P_{d,i,\max }},
\end{array}
\end{equation}
It is worth mentioning that after simplification, the QoS constraints in (\ref{eq:cell_opt_prob}) will be in terms of $x_{i}$ and hence the last two constraints in (\ref{eq:cell_opt_prob}) can be ignored at this problem.

The second derivative of $E{E_{c,i}}$ is given by,
\begin{equation}\label{eq:15}
\begin{array}{l}
\frac{{{\partial ^2}E{E_{c,i}}}}{{\partial P_{c,i}^2}}
= \frac{{2{W_i}\log_2\big(1 + {T_{c,i}}\big)\exp \big( { - {\varsigma _{c,i}}{\lambda _{c,i}}} \big)\exp \big( { - \frac{{{\varsigma _{d,i}}{\varsigma _{c,i}}{\lambda _{d,i}}{\lambda _{c,i}}}}{{\ln{x_i}}}} \big)}}{{P_{c,i}^3}}.
\end{array}
\end{equation}
It can be seen that $\frac{{{\partial ^2}E{E_{c,i}}}}{{\partial P_{c,i}^2}}$ is always positive. Thus, $E{E_{c,i}}$  is convex  everywhere. Hence, $E{E_c}$ which is the summation of convex functions is convex as well. Moreover, the constraints are linear functions of $P_{c,i}$ and thus the optimization problem is convex.

\subsection{Iterative Power Allocation}  

In previous subsections, we transformed the primary optimization problem into a convex form. However, these problems are not independent and should be solved jointly. This dependency is due to the impact of cellular users' transmission power on the entire network energy efficiency. To jointly solve the problem, first,  the  problem of D2D users' energy efficiency, with fixed transmission power of cellular users is solved. Afterward, using the solution of the D2D optimization problem, we optimize the transmission power of cellular users to maximize the network's energy efficiency. Since these problems must be solved consequently, we feed the result of cellular users' power to the D2D optimization problem in the next iteration, and we continue this procedure until the convergence criteria are met.

\begin{algorithm}[ht] 
	\caption{Iterative power allocation algorithm} \label{alg:1}
	\begin{algorithmic}[1] 
		\Statex	\textbf{Initialization}:
		\State \mbox{\boldmath$P_d$}$\leftarrow 0$, \mbox{\boldmath$P_c$}$\leftarrow 0$,  \mbox{\boldmath$\varepsilon_d$}$\leftarrow {10^{ - 5}}$, \mbox{\boldmath$\varepsilon_c$}$\leftarrow {10^{ - 5}}$, $n\leftarrow 0$, $N_{max} \leftarrow 10$ 
		\While{TRUE}{}
		\Statex \textbf{Phase I}
		\State 		Solve (\ref{eq:12}) for given \mbox{\boldmath$P_c$} and obtain \mbox{\boldmath$P_{d,cur}$} 
		\State \mbox{\boldmath$\Delta_d$}$=$\mbox{\boldmath$P_{d,cur}$}$-$\mbox{\boldmath$P_d$} and \mbox{\boldmath$P_d$}$=$\mbox{\boldmath$P_{d,cur}$} 
		\Statex \textbf{Phase II}
		\State Solve (\ref{eq:14}) for given \mbox{\boldmath$P_d$} and obtain \mbox{\boldmath$P_{c,cur}$}  
		\State \mbox{\boldmath$\Delta_c$}$=$\mbox{\boldmath$P_{c,cur}$}$-$\mbox{\boldmath$P_c$} and \mbox{\boldmath$P_c$}$=$\mbox{\boldmath$P_{c,cur}$} 
		\State  $n \leftarrow n+1$
		\If{$n==N_{max}$}
		\State Break
		\EndIf
		\If{ \mbox{\boldmath$\Delta_d$} $\le$ \mbox{\boldmath$\varepsilon_d$} and \mbox{\boldmath$\Delta_c$} $\le$ \mbox{\boldmath$\varepsilon_c$}}
		\State Break
		\EndIf
		\EndWhile
	\end{algorithmic}
\end{algorithm}

The proposed algorithm is reviewed in Algorithm \ref{alg:1}. The algorithm can be split into two phases. The first phase is D2D power allocation and the second one is cellular power allocation.  In the first phase, we initialize the fixed values of cellular users' parameters, e.g., transmission power, in $i$-th band and (\ref{eq:12}) is solved to find optimal power of D2D users in $i$-th band. Note that $x_i$ is convertible to $P_{d,i}$ in each iteration. In the second phase, the optimal power  of cellular users is achieved by utilizing the D2D users' power in the previous phase. In each iteration, ${\Delta _{c,i}}$ and ${\Delta _{d,i}}$ are computed, which are defined as ${\Delta _{c,i}} \buildrel \Delta \over = {{\rm{P}}_{{\rm{c,i,cur}}}}{\rm{ - }}{{\rm{P}}_{{\rm{c,i}}}}$ and ${\Delta _{d,i}} \buildrel \Delta \over = {{\rm{P}}_{{\rm{d,i,cur}}}}{\rm{ - }}{{\rm{P}}_{{\rm{d,i}}}}$. ${P_{c,i,cur}}$, ${P_{d,i,cur}}$ and ${P_{c,i}}$, ${P_{d,i}}$ are the optimum power of cellular and D2D users in $i$-th band in the current iteration and previous iteration, respectively. The iteration  will continue until ${\Delta _{c,i}} \le {\varepsilon _{c,i}}$ and ${\Delta _{d,i}} \le {\varepsilon _{d,i}}$  or the maximum iteration number is reached which is denoted by $N_{max}$.  ${\varepsilon _{d,i}}$  and ${\varepsilon _{c,i}}$ are the maximum tolerance of D2D users' power and cellular users' power in $i$-th band, respectively. According to \cite{zhou2016energy}, the iterative optimization algorithm converges to the maximum EE if problems (\ref{eq:12}) and (\ref{eq:14}) are feasible. For obtaining optimum power in each iteration, the interior point method which is a numerical solver is used. For this purpose, we use OPTI toolbox which is a free MATLAB toolbox for optimization  \cite{19}.

\subsection{Complexity Analysis}
In this subsection, we investigate the computational complexity of the proposed power allocation algorithm. In this algorithm, after initialization, we have to solve the optimization problem \eqref{eq:12}. In this problem, in total we have $M$ number of decision variables and $3M+1$ linear and convex constraints where $M$ is total number of bands. Hence, the computational complexity of solving this problem is $O\big(M^3 (3M+1)\big)$ \cite{masoudi2017cloud}.
Similar to this approach, solving  \eqref{eq:14} has the computational complexity of $ O\big(M^3 (2M+1)\big) $. Other operations in the loop of power allocation algorithm have the computational complexity of $O\big(1\big)$ and can be ignored compared to complexity order of solving optimization problems. Therefore, assuming that,$N_{max}$ is the maximum number of loop iterations, the computational complexity of proposed algorithm is given by
\begin{eqnarray}
O\big([M^3 (3M+1) + M^3 (2M+1)]\big) \approx O\big(N_{max} M^4\big)
\end{eqnarray}

\section{Numerical Results}
In this section,  the energy efficiency of the   D2D  and cellular users, namely  $EE_{c}$ and $EE_{d}$ is evaluated. According to  Fig. \ref{fig:sys}, we consider an uplink system where both cellular users and devices share the same spectrum with $M$ different bands. The simulation parameters are shown in Table 1, and since we want to compare our results with a benchmark, we kept the simulation parameters as is set in \cite{15}.

\begin{figure}[ht]
	\centering
	\includegraphics[width=0.5\linewidth]{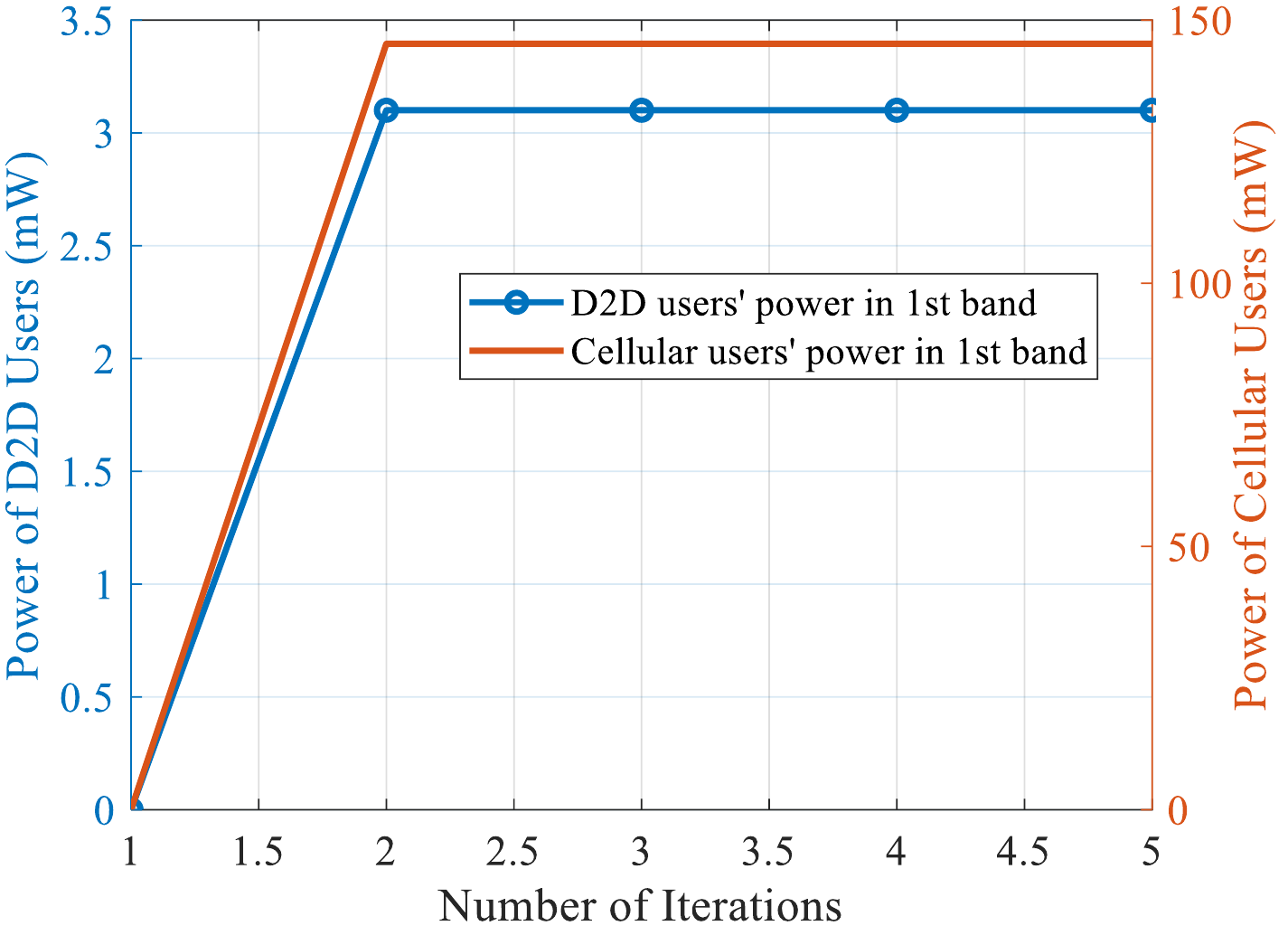}
	\caption{Power of users against number of iterations.}
	\label{fig:it}
\end{figure}

Fig. \ref{fig:it} represents the convergence behavior of Algorithm \ref{alg:1}, for the transmission power of cellular and D2D users on one band with  ${\lambda _{d,ref}}=\, {10^{ - 4}}$ and ${\lambda _{c,ref}}=\, {15 \times 10^{ - 6}}$.  Fig. \ref{fig:it} demonstrates that  Algorithm \ref{alg:1} converges to the optimum value within a few iterations.  According to (\ref{eq:5}), the capacity is averaged over the distribution of channel so the effect of channel randomness is eliminated and due to less variation of channel, very few iterations are needed for our algorithm to be converged.

\begin{table}[ht]
	\centering
	\caption{Simulation parameter values}
	\label{tab:1}       
	\begin{tabular}{lll}
		\hline\noalign{\smallskip}
		Parameter & Value  \\
		\noalign{\smallskip}\hline\noalign{\smallskip}
		$M$  & 5 \\
		$W_{i}$	 & 20 MHz \\
		$\alpha $ & 4\\
		${\theta _{c,i}} $ & 0.05 \\
		${\theta _{d,i}} $ & 0.05 \\
		$\left[{{R_{d,00,1}},{R_{d,00,2}},...,{R_{d,00,5}}} \right] $ & $\left[ {10,\,20,\,30,\,20,\,10} \right]\,m$ \\
		$\left[{{R_{c,00,1}},{R_{c,00,2}},...,{R_{c,00,5}}} \right] $  &  $\left[ {50,\,60,\,70,\,80,\,90} \right]\,m$ \\
		$\left[ {{\lambda _{d,1}},{\lambda _{d,2}},...,{\lambda _{d,5}}} \right] $  & $\left[ {10,\,1,\,10,\,10,\,10} \right]\, \times {\lambda _{d,ref}}$ \\
		$\left[ {{\lambda _{c,1}},{\lambda _{c,2}},...,{\lambda _{c,5}}} \right] $ & $\left[ {10,\,1,\,10,\,10,\,10} \right]\, \times {\lambda _{c,ref}}$ \\
		${P_{d,i,\max }} $  & 20 mW \\
		${P_{c,i,max}} $  & 300 mW \\
		${P_c} $ & 1 W \\
		${\varepsilon _{d,i}} $ & ${10^{ - 5}}$ \\
		${\varepsilon _{c,i}} $ & ${10^{ - 5}}$ \\
		\noalign{\smallskip}\hline
	\end{tabular}
\end{table}

\begin{figure}[ht]
	\centering
	\includegraphics[width=0.5\linewidth]{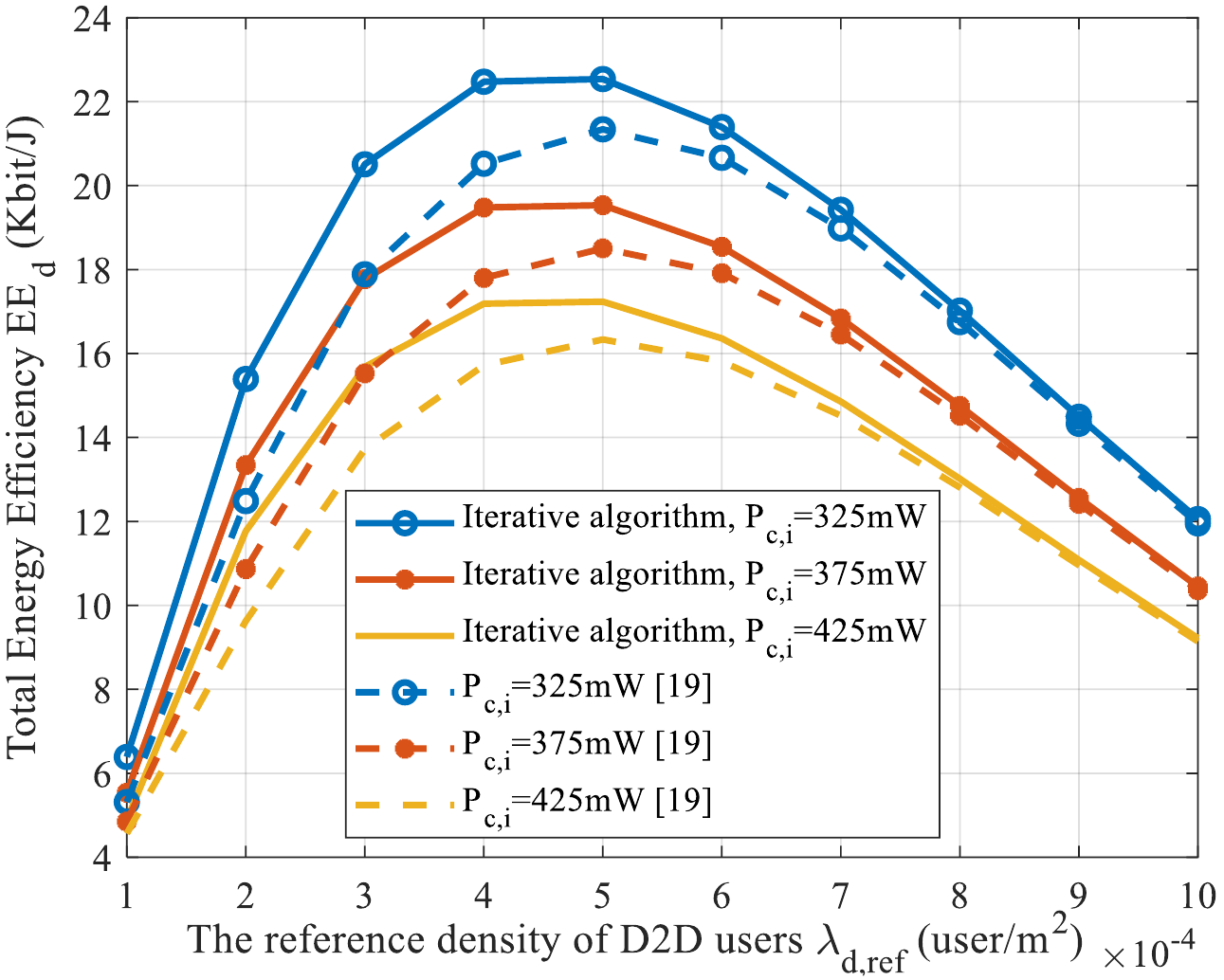}
	\caption{Energy efficiency of D2D users versus ${\lambda _{d,ref}}$. }
	\label{fig:EE_D2D_D}
\end{figure}

\begin{figure}[ht]
	\centering
	\includegraphics[width=0.5\linewidth]{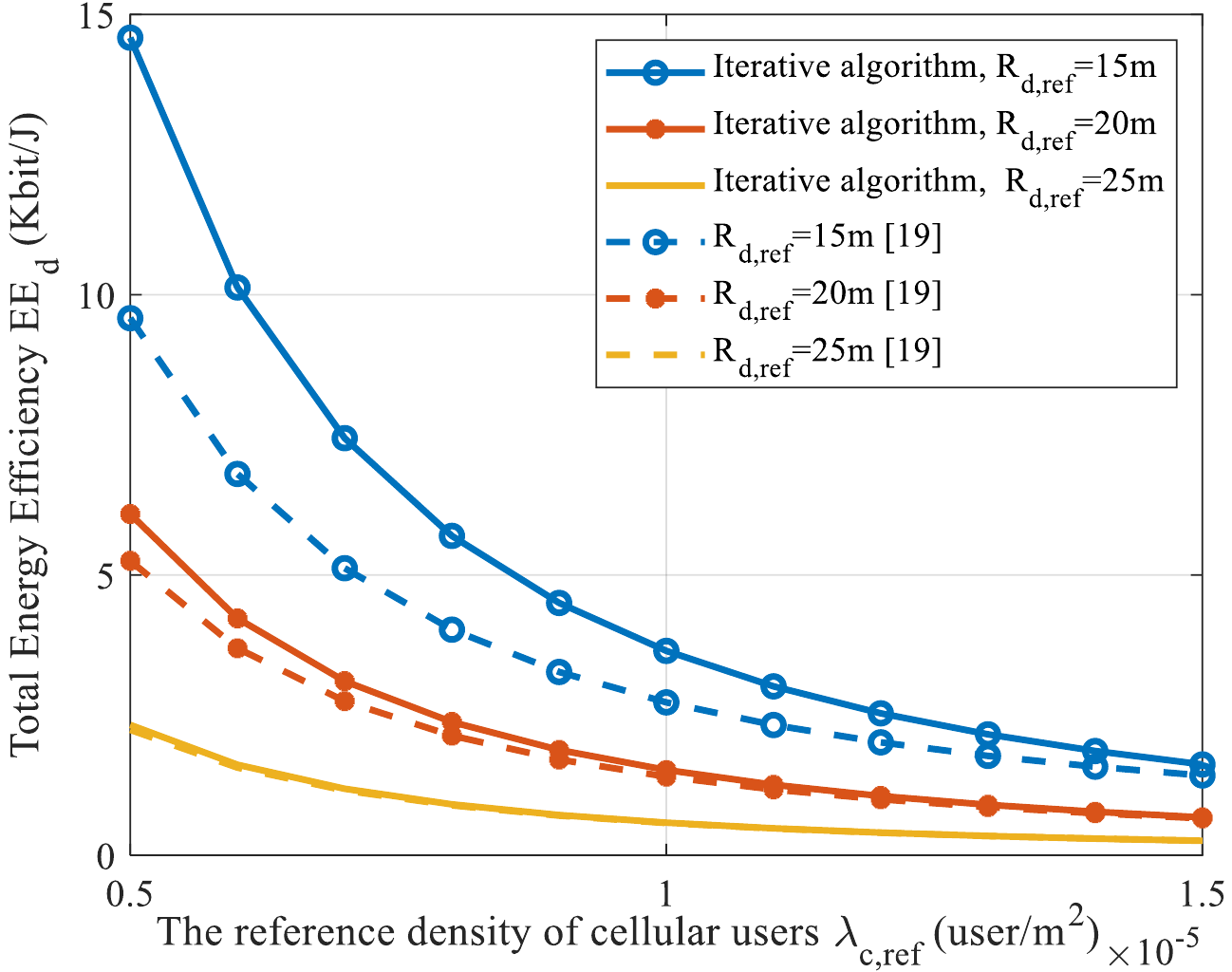}
	\caption{Energy efficiency of D2D users versus ${\lambda _{c,ref}}$. 	}
	\label{fig:EE_D2D_C}
\end{figure}

In Fig. \ref{fig:EE_D2D_D} and \ref{fig:EE_D2D_C}, we investigated the EE of D2D users with regards to the density of D2D and cellular users. In these figures, we have solved the optimization problem in (\ref{eq:12}), and the optimal power of D2D users in each band is derived by OPTI toolbox. As a benchmark, we compared our results with  \cite{15}, where the problem of D2D energy efficiency in (\ref{eq:11}) is solved  with a derivative-based algorithm. It is worth mentioning that in their approach they did not consider cellular users' power and hence to compare our results; first, we fix the transmission power of cellular users.  Fig. \ref{fig:EE_D2D_D} and \ref{fig:EE_D2D_C} reveals that in terms of EE, Algorithm \ref{alg:1} with fixed cellular users power transmission, outperforms the algorithm in \cite{15}. Later, we improve the EE even more by enhancing cellular users' transmission power. 

In Fig. \ref{fig:EE_D2D_D}, total  power of devices in all bands, $P_d$, is set to $60$ mW  and ${\lambda _{c,ref}}$ is ${10^{ - 5}}$. It can be seen that by increasing ${\lambda _{d,ref}}$, EE increases at first and then declines. The STP increases as the density of D2D users become sparser; thus ASR of D2D users increases according to (\ref{eq:5}), and EE increases too. In higher densities, the interference which is produced by spectrum sharing is more than ASR of D2D users. So, the EE decreases. Also, by increasing the cellular users' transmission power, the produced interference by cellular users increases and EE decreases.

In Fig. \ref{fig:EE_D2D_C}, the simulation parameters are set as: $\left[ {{R_{d,00,1}},\,{R_{d,00,2}},\,{R_{d,00,3}},\,{R_{d,00,4}},\,{R_{d,00,5}}} \right]\, = \,{R_{d,ref}}  \times  \left[ {1,2,3,2,1} \right]$,  $P_d=\,80\,mW$,  ${\lambda _{d,ref}}=\, {10^{ - 4}}$ and $P_{c,i}=\,300\,mW$.  As can be seen in Fig. \ref{fig:EE_D2D_C}, when ${\lambda _{c,ref}}$ increases, the interference of cellular users on devices increases and devices should use more power to coordinate the interference. So, EE decreases. Also, by increasing the distance between D2D users, EE decreases because the channel fading becomes greater. 

\begin{figure}[ht] 
	\centering
	\includegraphics[width=0.5\linewidth]{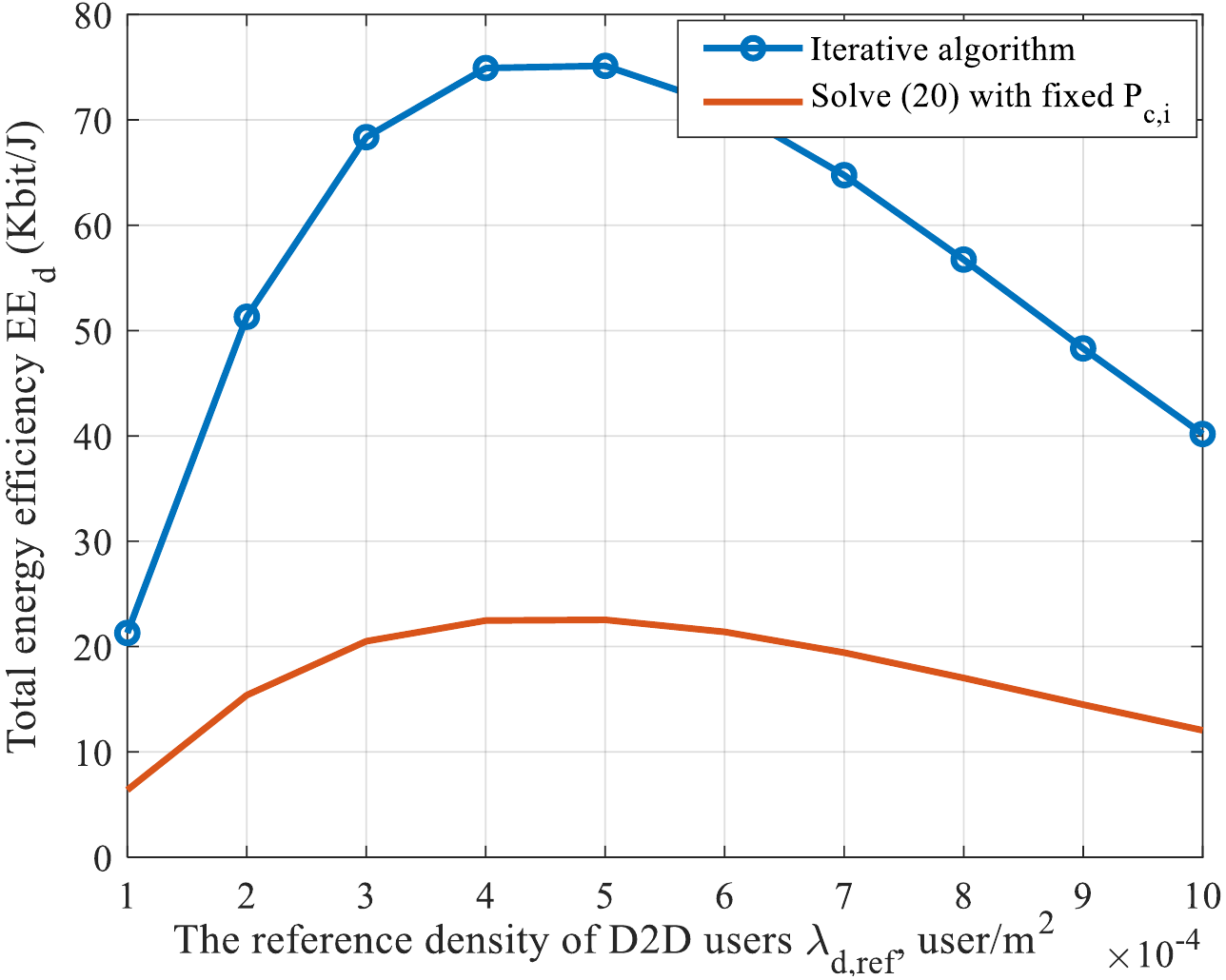}
	\caption{Energy efficiency of D2D users versus ${\lambda _{d,ref}}$ by obtaining power of cellular and D2D users, iteratively. 
	}\label{EEd_OP_P}
\end{figure}

The  cellular users' transmission power has an impact on the EE of devices. This impact is investigated in the following results. The comparison benchmark is the case when (\ref{eq:12}) is solved numerically and the cellular users' transmission power  is fixed  as $325\, mW$.  In Fig. \ref{EEd_OP_P}, we have depicted $EE_{d}$ versus  density of devices. Fig. \ref{EEd_OP_P} shows that optimizing the transmission power of cellular users and devices jointly can significantly improve the $EE_d$ since the destructive impact of cellular users on the D2D users is now mitigated.

\begin{figure}[ht] 
	\centering
	\includegraphics[width=.5\linewidth]{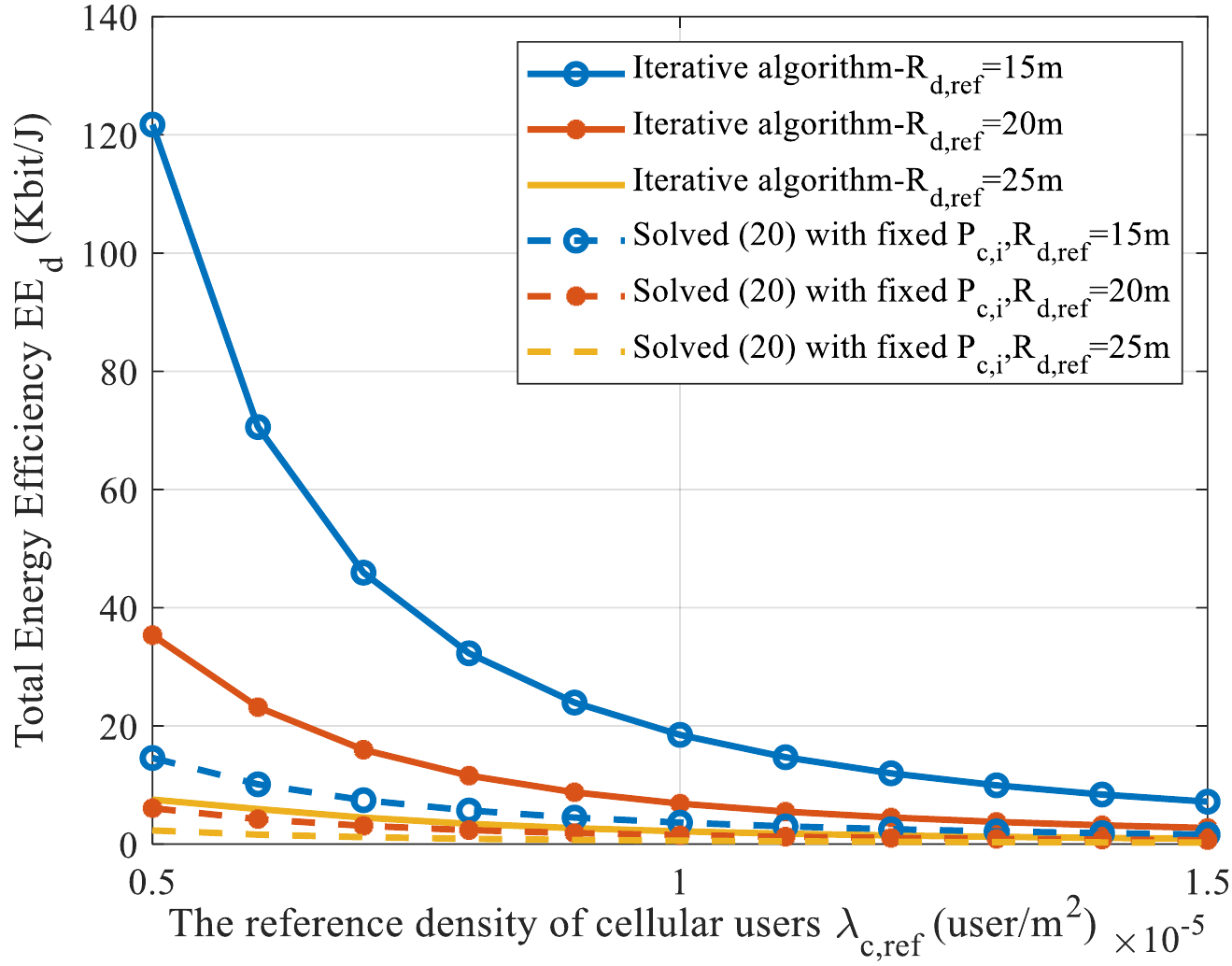}
	\caption{Energy efficiency  of D2D users versus ${\lambda _{c,ref}}$ by obtaining power of cellular users and devices, iteratively. 
	}\label{EEd_OP_C}
\end{figure}

In Fig. \ref{EEd_OP_C}, we investigated the impact of cellular users density on  $EE_d$. As the density of cellular users increases, $EE_d$ decreases since cellular users create more interference on D2D users. To keep the link quality, D2D users should consume more power and as a result, $EE_d$ decreases. Algorithm \ref{alg:1} still outperforms the reference scenario; however, at very dense scenario the gap is low since there is no room for enhancing the $EE_d$. Moreover, we plotted the $EE_d$ for different $R_{d,ref}$, which is defined as, the distance between a pair of D2D connection. It can be seen that higher  $R_{d,ref}$ results in lower $EE_d$ since  this further distance means that more power is required to meet QoS requirements.
Finally, as can be seen in Fig. \ref{EEd_OP_P} and Fig. \ref{EEd_OP_C}, $EE_d$ is higher than $EE_d$ in Fig. \ref{fig:EE_D2D_D} and Fig. \ref{fig:EE_D2D_C} and this highlights the impact of optimizing the cellular users' transmission power.

\begin{figure}[ht] 
	\centering
	\includegraphics[width=0.5\linewidth]{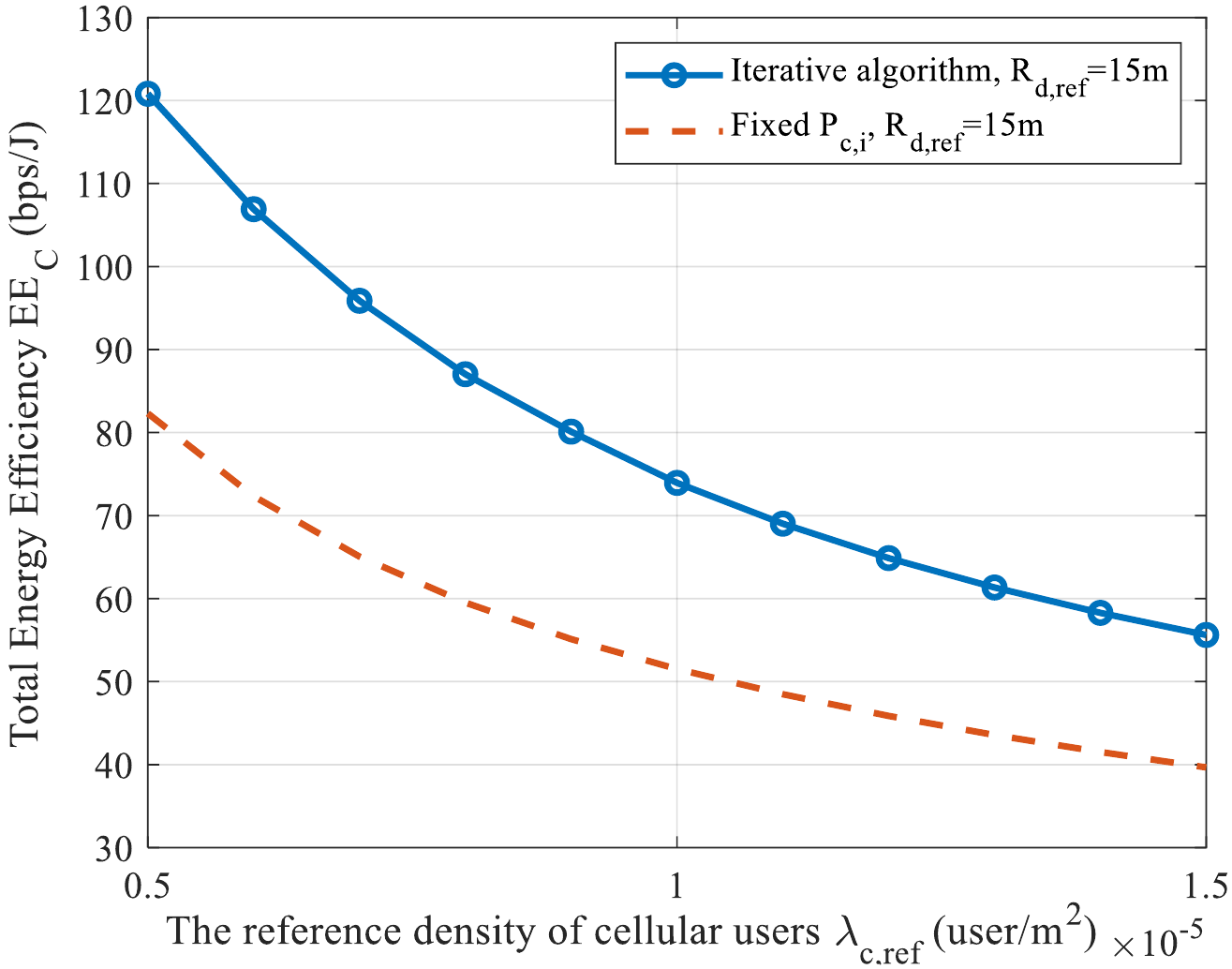}
	\caption{Energy efficiency of cellular users versus ${\lambda _{c,ref}}$ by obtaining power of cellular users and devices, iteratively. 
	}\label{EEd_O}
\end{figure}

In Fig. \ref{EEd_O}, EE of cellular users, $EE_{c}$,  versus ${\lambda _{c,ref}}$ is illustrated. We compare the $EE_{c}$ of Algorithm \ref{alg:1} with the $EE_{c}$ of \cite{15}. To calculate the $EE_c$ in \cite{15},  we first solved (\ref{eq:12}) to achieve the  D2D users' transmission power. Then, assuming that the cellular users' transmission power is constant, $P_{c,i} = 325$\, $mW$,  we substituted the  power of devices in  (\ref{eq:10}), and $EE_{c}$ can be calculated accordingly. By increasing ${\lambda_{c,ref}}$, the interference created by cellular users increases and hence $EE_c$  decreases.  As can be seen in Fig. \ref{EEd_O}, we can achieve higher energy efficiency by Algorithm \ref{alg:1} because the cellular users are operating with their optimal transmission power.

\section{Conclusions}
In this study,  the EE maximization of the D2D communication underlaying cellular networks is investigated. First, we derived the closed-form expressions for the successful transmission probability (STP) and average sum rate (ASR) of cellular and D2D users. Then, we have formulated an optimization problem that maximizes the energy efficiency of both cellular and D2D users as MINLP problem. In the optimization problem, we considered STP as QoS of cellular and D2D users. To solve the optimization problem, first, we converted the primary optimization problem into a  canonical convex form. Then, we break down the problem into two sub-problems. The first sub-problem was devoted to solving the problem of D2D users energy efficiency.  Utilizing the result of the first sub-problem, we maximized the energy efficiency of cellular users in the second sub-problem. Our simulation results  showed that  our scheme outperforms the existing solutions in the literature. Moreover, significant enhancement in terms of energy efficiency is achieved considering optimizing D2D and cellular users transmission power jointly.  As a future work,  band selection can be considered along with power allocation to minimize the cross-tier interference between D2D users and cellular users. Moreover, the optimal density of D2D users and cellular in each band and the corresponding trade-off analysis can be investigated.

\ifCLASSOPTIONcaptionsoff
  \newpage
\fi





\bibliographystyle{ieeetr}
\bibliography{refs}

\end{document}